\renewcommand{\baselinestretch}{1}

\documentclass[9pt, conference]{IEEEtran}
\usepackage{balance}
\usepackage{float} 
\usepackage{tabularx}
\usepackage{mathtools}
\usepackage{amsmath}
\usepackage{hhline}
\usepackage{hyperref}
\usepackage{silence}
\usepackage{graphicx}
\usepackage{multirow} 
\usepackage{graphicx}
\usepackage{multirow}
\usepackage{amsmath}
\usepackage[english]{babel}
\usepackage{booktabs}
\usepackage{array}
\usepackage{paralist}
\usepackage{threeparttable}
\usepackage{lipsum}
\usepackage{flushend}
\usepackage{cuted}
\usepackage{algpseudocode}
\usepackage{algorithm}
\usepackage{algcompatible}
\usepackage{amssymb}
\usepackage{color}
\usepackage{psfrag}
\usepackage{epsfig}
\usepackage{multicol}
\usepackage{color}
\usepackage{fancybox}
\usepackage{subfigure}
\usepackage{cite}
\usepackage{theorem}
\usepackage{url}
\usepackage[table]{xcolor}
\usepackage{epstopdf}
\usepackage[normalem]{ulem}
\usepackage{siunitx}
\usepackage{rotating}

\DeclareUnicodeCharacter{2009}{\,}

\usepackage[justification=centering,font=small,labelfont=bf]{caption}
\usepackage{silence}
\usepackage{enumitem}
\definecolor{gray1}{gray}{0.90}
\definecolor{gray2}{gray}{0.98}
\definecolor{light-gray}{gray}{0.95}

\newcommand{\ignore}[1]{}

\usepackage{tikz}
\usetikzlibrary{tikzmark}
\renewcommand{\baselinestretch}{0.93}
\begin{document}

\title{A Life-Cycle Energy and Inventory Analysis of Adiabatic Quantum-Flux-Parametron Circuits}
\author{
Masoud Zabihi, Yanyue Xie, Zhengang Li, Peiyan Dong, Geng Yuan, Olivia Chen, Massoud Pedram, Yanzhi Wang\\
%Northeastern University, 360 Huntington Ave, Boston, MA 02115
}

\maketitle
\begin{abstract}
The production process of superconductive integrated circuits is complex and consumes significant amounts of resources and energy. Therefore, it is crucial to evaluate the environmental impact of this emerging technology. An attractive option for the next generation of superconductive technology is Adiabatic Quantum-Flux-Parametron (AQFP) devices. This study is the first to present a comprehensive process-based life-cycle assessment (LCA) and inventory analysis of AQFP integrated circuits. To generate relevant outcomes, we conduct a comparative LCA that included the bulk CMOS technology. The inventory analysis considered the manufacturing, assembly, and use phases of the circuits. To ensure a fair assessment, we choose the 32-bit AQFP RISC-V single-core processor as the reference functional unit and compare its performance with that of a CMOS counterpart. Our findings reveal that the AQFP processor consumes several orders of magnitude less energy during the use phase than its CMOS counterpart. Consequently, the total life cycle energy (which encompasses manufacturing and assembly energies) of AQFP integrated circuits improves at least by two orders of magnitude.

\makeatletter
\renewcommand*{\@makefnmark}{}
\makeatother
\end{abstract}
%\makeatletter{\renewcommand*{\@makefnmark}{}
%\footnotetext{* (equal contribution)}\makeatother}

\section{Introduction} \label{sec:intro}
%%% Rewriting %%%%
%\noindent

Compared to traditional CMOS technology, superconductive electronic circuits (and particularly AQFP-based logic) have advantages such as significantly lower power consumption~\cite{superconductive_advantage}. It is forecasted that AQFP-based logic can reach tens of thousands of times energy-per-operation advantages over state-of-the-art CMOS counterparts. As a result, AQFP superconductive electronics have the potential to revolutionize supercomputing systems by reducing energy consumption and improving performance~\cite{Olivia_paper}.

Evaluating the resources and energy of emerging circuits during manufacturing, assembly, and use phases is a vital part of environmental impact mitigation, and LCA is increasingly used for this purpose. While LCA for CMOS circuits has been thoroughly researched~\cite{nature_LCA}, there is still much to be learned about the environmental impact of AQFP circuits. Performing a LCA for both CMOS and AQFP circuits is crucial to determine which technology is superior overall and to ensure the industry's sustainable and efficient use of resources.

This paper presents a comprehensive life-cycle energy and inventory analysis of AQFP superconductive integrated circuits, which are compared to CMOS technology circuits. The analysis encompasses the manufacturing, assembly, and use phases, utilizing 32-bit processors based on RISC-V architecture for both AQFP and CMOS circuits as the functional unit. The contribution of the paper is multifold: (i) presenting the first LCA of the AQFP circuit and including key parameters such as cooling cost in the LCA flow, (ii) analyzing the manufacturing steps of the AQFP wafer and providing comparisons with CMOS counterparts, (iii) yield analysis for both technologies, and examining the manufacturing, assembly, and use phase energies of the RISC-V AQFP-based processor and its CMOS counterpart, (v) investigation of the downscaling effect of the AQFP chip area on the LCA.

This paper is structured as follows: a description of the LCA method is provided in Section II, followed by a discussion of the results for life-cycle energy and inventory analysis of AQFP integrated circuits in Section III. The conclusion is presented in Section IV.

%%%% The comparison table
\begin{table*}[!ht]
\renewcommand{\arraystretch}{1.3}
\caption{Comparison of CMOS and AQFP RISC-V Processors}
\label{table:comparison}
\centering
\begin{tabular}{c||c|c|c|c||c}
\hline
Processor & Manufacturing Energy & Assembly Energy & Use Phase Energy & Total Energy & Overall Improvement \\ 
\hline\hline
CMOS RISC-V & 0.17 KWh & 0.08 KWh & 665.23 KWh & 665.48 KWh & \\ 
\cline{1-5}
\raisebox{1ex}{AQFP RISC-V} & \raisebox{1ex}{1.61 KWh} & \raisebox{1ex}{1.19 KWh} & \shortstack{\\[0.1pt]0.001 KWh\\(with cooling 0.42 KWh)} & \shortstack{2.81 KWh\\(with cooling 3.23 KWh)} & \raisebox{2ex} {\shortstack{237X\\(with cooling 205X)}} \\ 
\hline
\end{tabular}
\end{table*}

\section{LCA overview and Method}
\label{sec:background}

 In this work, we follow a process-based LCA due to its consistency and ability to provide provides a more detailed and specific understanding of the life cycle. Other than process-based LCA, economic input-output LCA is the most common approach and a valuable tool for determining the impact of economic activities across sectors~\cite{EIO_LCA}. LCA considers the entire life cycle of a product, from raw material extraction to end-of-life disposal, in order to identify and address environmental impacts. One of the key aspects of LCA is defining the system boundaries, which determine which stages of the life cycle are included in the assessment. These boundaries can be narrow, focusing on specific parts of the product's life cycle, or broad, encompassing the entire supply chain. 

\begin{figure}[t!]
  \centering
  \includegraphics[width=0.5\textwidth]{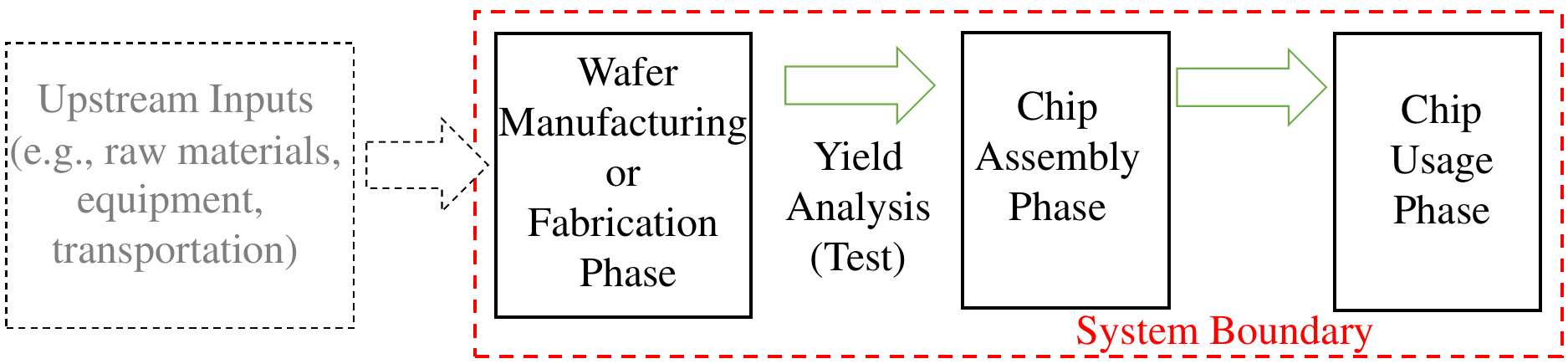}
  \caption{ Scope and boundary of analysis for CMOS or AQFP circuits.}
  \label{fig:LCA_boundry}
\end{figure}

Figure~\ref{fig:LCA_boundry} illustrates the boundaries and extent of our study. We present a thorough analysis of the inventory and energy, encompassing the manufacturing, assembly, and use phases of both CMOS and AQFP integrated circuits. The accounting of upstream inputs (e.g., raw materials extraction and transportation) has not been considered due to the high degree of uncertainty involved. Our investigation provides a detailed account of the life-cycle inventory and energy analysis which includes the following aspects: functional unit selection, the wafer manufacturing and fabrication phase, wafer cutting and yield analysis, assembly phase, and use phase, which will be discussed in the rest of this section. 

A. Functional unit selection: The life-cycle impacts of CMOS/AQFP circuits can be significantly influenced by the selection of the functional unit. To perform an ``apple-to-apple'' comparison and ensure the same level of functionality, we choose the single-core 32-bit CMOS RISC-V architecture for both CMOS and AQFP technologies as the functional unit. The AQFP RISC-V processor includes crucial components such as a decoder, 32-bit ALU, register file, controller, and L1 AQFP cache memory~\cite{AQFP_memory}. The total area of the AQFP RISC-V processor is obtained by the summation over component areas. The synthesis of the AQFP component circuits is based on the MIT-LL process~\cite{MITLL_process}.
Based on our synthesis, the clock frequency, power, and area for the AQFP processor are 5 GHz, \SI{41}{\micro\watt}, and 3.5~cm$^2$, respectively. Note that to ensure a fair evaluation, we compare today's RISC-V AQFP processors built out of micron-size AQFP devices with a 130 nm CMOS processor that has its fabrication process steps provided in~\cite{fabrication_steps}. Moreover, note that for state-of-the-art technology nodes (e.g., 7nm), the full process steps details are not available. We used Western Digital SweRV EH1 RISC-V processor features as a sample and used scaling equations provided in~\cite{scalling_formulas} to obtain the equivalent 130 nm chip features. For CMOS, the clock frequency, power, and area are 1 GHz, \SI{7.5}{\watt}, and 12.1~mm$^2$, respectively.

B. The manufacturing processes for AQFP and CMOS circuits share some similarities, but there are also notable differences in materials and specific steps. AQFP technology typically employs simpler steps than those used in semiconductor fabrication plants. For CMOS, a complete inventory of fabrication steps and their associated energy requirements for a 300 mm wafer can be found in Reference~\cite{fabrication_steps}, where a total of 206 process steps are listed in Tables S3 to S17 of the supplementary materials. To evaluate the fabrication process for AQFP technology, we carefully reviewed the process presented in~\cite{Japan_process_steps}, and estimated a total of 216 steps and their corresponding energy values. The total energy required for each wafer is obtained by summing the energy values of all steps.

C. Wafer cutting: After determining the energy required for fabricating a wafer calculate the fabrication yield. The yield for CMOS wafers is found (using Murphy’s model) to be 97.6\%, while the yield for AQFP wafers was estimated to be 85.2\%. Using yield, we calculate the number of functional dies for each technology. The energy for the fabrication of each die is calculated using the formula: Manufacturing Energy of Wafer / Number of Functional Dies. 

D. Assembly phase: In our analysis, we consider the energy required for the packaging and assembly of each die. We consider commonly used plastic packages. The energy usage in the packaging stage is 0.34kWh per cm\textsuperscript{2} of silicon~\cite{packaging_energy}.

E. Use phase: We assume the usage as the server for both technologies. AQFP and CMOS circuit lifetimes are considered to be 10 and 5 years, respectively. Large computing systems
require cooling. This is particularly important for superconducting electronics which operate at temperatures below 10 K. We consider the cooling energy cost of the superconducting circuit to be 400X larger than the energy generated by the circuit at the cryogenic temperature~\cite{400X_cooling}.

\section{Results and Discussions}
\label{sec:ir-drop-similarity}

Table~\ref{table:comparison} presents a concise summary of the life-cycle energy consumption results for AQFP and CMOS technologies. Notably, the cooling cost of AQFP circuits is taken into account in our report. In the case of CMOS technology, the dominant energy component is the use of phase energy, which is considerably greater than manufacturing and assembly energies. However, for AQFP technology, manufacturing and assembly energies are significantly higher than the use phase energy.

A comparison of the energy components of the two technologies reveals that the manufacturing and assembly energies for AQFP processors are respectively 9.5X and 14.8X larger than those for a CMOS processor. This disparity can be attributed to the fact that the AQFP chip has a 28.9X larger area than that of CMOS. On the other hand, AQFP technology is significantly more energy-efficient than CMOS technology during the use phase, even when the cooling effect is taken into account. Without cooling cost, AQFP technology is six orders of magnitude more energy-efficient than CMOS during the use phase, and with cooling cost, it is four orders of magnitude more energy-efficient. The AQFP technology's overall superiority over CMOS technology is primarily due to its extreme energy efficiency during the use phase, which significantly outperforms the CMOS counterpart. Overall, considering all phases, AQFP processor is 237X and 205X more energy-efficient than CMOS processor in with and without considering the cooling effect, respectively. . 

In comparison to CMOC, AQFP technology is still in its nascent stages. It's important to note that for this assessment, we utilized today's micron-size AQFP technology against 130 nm CMOS technology node. However, Our investigation suggests that reducing the size of AQFP chips in the near future is inevitable, for example, due to advancements in AQFP device technology and more efficient, area-aware routing algorithms. Even a small downscaling (e.g., 2X) of the AQFP chip area can result in a significant improvement in fabrication yield and overall life cycle energy.  Our findings indicate that a 2X downscaling of AQFP chip area can improve wafer yield from 85.2\% in today's AQFP technology to 92.1\% in the near future AQFP technology. Moreover, we anticipate that the combined manufacturing and assembly energy requirements will decrease by 52\%, from a total of 2.8 KWh in today's AQFP technology to 1.3 KWh in the near future AQFP technology. This is expected to result in an overall improvement (compared to CMOS) in life cycle energy efficiencies of 498X and 378X without and with cooling costs, respectively.

%%%% The comparison table including RRAM technology
\ignore{
\begin{table*}[!ht]
\renewcommand{\arraystretch}{1.3}
\caption{Comparison of Energy Efficiency Between CMOS and AQFP RISC-V Processors for General Purpose Computing and RRAM/AQFP Crossbars for Special Purpose Computing.}
\label{table:comparison}
\centering
\begin{tabular}{c|c||c|c|c|c||c}
\hline
Application & Technology & Manufacturing Energy & Assembly Energy & Use Phase Energy & Total Energy & Improvement \\ 
\hline\hline
 \raisebox{-2ex} {GPC} & CMOS CPU & 0.79 KWh & 0.34 KWh & 722.70 KWh & 1,447.67 KWh &  \\ 
\cline{2-6}
 &\raisebox{1ex}{AQFP RISC-V } & \raisebox{1ex}{5.94 KWh} & \raisebox{1ex}{1.67 KWh} & \shortstack{\\[0.1pt]0.034 KWh\\(with cooling 3.48 KWh)} & \shortstack{7.61 KWh\\(with cooling 11.06 KWh)} & \raisebox{2ex} {\shortstack{190X\\(with cooling 130X)}} \\ 
\hline\hline
\raisebox{-2ex} {SPC} & RRAM crossbar & 7.05e-5 KWh & 3.48e-5 KWh &  25.68 KWh & 51.37 KWh &  \\ 
\cline{2-6}
 & \raisebox{1ex}{AQFP crossbar} & \raisebox{1ex}{2.91 KWh} & \raisebox{1ex}{ 1.43 KWh} & \shortstack{\\[0.1pt]5.60e-4 KWh\\(with cooling 0.22 KWh)} & \shortstack{4.35 KWh\\(with cooling 4.57 KWh)} & \raisebox{2ex} {\shortstack{12 X\\(with cooling 11X)}} \\ 
\hline

\end{tabular}
\end{table*}
}

\noindent
\section{Conclusion}
\label{sec:conclusion}
In this paper, we describe the first attempt at a comparative
life-cycle energy and inventory analysis between the AQFP integrated circuits and those of a conventional
CMOS technology. We provide energy and inventory
analysis accounting for the manufacturing, assembly, and use phases. The functional units used in this paper are 32-bit single-core CMOS and AQFP RISC-V processors. Our analysis shows that while the AQFP processor has larger manufacturing and assembly energy, the use phase energy consumed by AQFP circuits is significantly smaller than the CMOS counterpart. As a result, AQFP circuits are 237X more energy efficient in total in comparison with the CMOS counterpart. Even with considering the cooling cost the energy efficiency of the AQFP circuit is 205X. In the near future, downsizing the AQFP chip area can lead to a significant reduction in the manufacturing and assembly energy of the AQFP circuits. Moreover, the overall life cycle of the AQFP circuits can be improved several times over through such downsizing.   
\noindent

\bibliographystyle{misc/ieeetr2}
\bibliography{misc/bibfile}
%misc/cram,misc/main.bib,misc/pim.bib}

%\input{sec/9-appendix}

\clearpage

\end{document}